\documentclass[journal=nalefd,manuscript=review]{achemso}

\usepackage[version=3]{mhchem} 

\author{D. Gustafsson}
\affiliation[Microtechnology and Nanoscience] {Department of
Microtechnology and Nanoscience, Chalmers University of Technology,
SE-412 96 Göteborg, Sweden}\email{dagust@chalmers.se}
\author{H. Pettersson}
\affiliation[Microscopy and Microanalysis]{Microscopy and
Microanalysis, Department of Applied Physics, Chalmers University of
Technology, SE-412 96 Göteborg, Sweden}
\author{B. Iandolo}
\affiliation[Microtechnology and Nanoscience] {Department of
Microtechnology and Nanoscience, Chalmers University of Technology,
SE-412 96 Göteborg, Sweden}
\author{E. Olsson}
\affiliation[Microscopy and Microanalysis]{Microscopy and
Microanalysis, Department of Applied Physics, Chalmers University of
Technology, SE-412 96 Göteborg, Sweden}
\author{T. Bauch}
\affiliation[Microtechnology and Nanoscience] {Department of
Microtechnology and Nanoscience, Chalmers University of Technology,
SE-412 96 Göteborg, Sweden}
\author{F. Lombardi}
\affiliation[Microtechnology and Nanoscience] {Department of
Microtechnology and Nanoscience, Chalmers University of Technology,
SE-412 96 Göteborg, Sweden}

\title[Soft nanostructuring of YBCO Josephson Junctions by phase separation]
{Soft nanostructuring of YBCO Josephson Junctions by phase
separation}

\begin{document}

\begin{abstract}
We have developed a new method to fabricate biepitaxial
YBa$_2$Cu$_3$O$_{7-\delta}$ (YBCO) Josephson junctions at the
nanoscale, allowing junctions widths down to 100 nm and
simultaneously avoiding the typical damage in grain boundary
interfaces due to conventional patterning procedures. By using the
competition between the superconducting YBCO and the insulating
Y$_2$BaCuO$_5$ phases during film growth, we formed nanometer sized
grain boundary junctions in the insulating Y$_2$BaCuO$_5$ matrix as
confirmed by high resolution transmission electron microscopy.
Electrical transport measurements give clear indications that we are
close to probing the intrinsic properties of the grain boundaries.
\end{abstract}
\textit{This document is the Accepted Manuscript version of a
Published Work that appeared in final form in Nano Letters,
copyright \copyright American Chemical Society after peer review and
technical editing by the publisher. To access the final edited and
published work see
\newline http://pubs.acs.org/doi/full/10.1021/nl103311a}
\newline
\newline
With the advances in nanotechnologies it has become important to
develop quantum limited sensors which allow to measure the physics
of very small objects with nanometer size resolution. Josephson
junction based nanodevices, as an example nanoSQUIDs
(Superconducting Quantum Interference Device), are very attractive
for their extreme sensitivity to magnetic flux which should allow as
ultimate goal the detection of individual spin\cite{Lam}, for their
potentiality in quantum informatics processing based on spin
systems\cite{Meier} and for their employment in scanning SQUID
microscopy\cite{Foley}. The miniaturization of the Josephson
junctions is the key issue in this area. Recent advances have
prioritized simple fabrication procedures which have led to the use
of nanobridge junctions instead of the commonly used Josephson
tunnel junctions. In this way planar Al and Nb nanoSQUIDs with
excellent noise performances have been fabricated with a single
layer superconductive film by e-beam lithography\cite{Lam} or
Focused Ion beam (FIB)\cite{Hilghenkamp}. The assessment of single
layer technology for nanoscale high-temperature superconductor (HTS)
Josephson junctions would certainly push this field forward. HTS
nanoSQUIDs would allow to study magnetic nanoparticles at much
higher magnetic fields when compared to low critical temperature
superconductors (LTS).

The miniaturization of HTS oxides has followed different
nanofabrication routes compared to the LTS counterpart. There are
two main reasons: a) The high temperature involved in the epitaxial
growth of thin films ($\sim$ 700$^\circ$C) makes standard lift- off
patterning procedures not applicable and b) conventional Reactive
Ion Etching (RIE) techniques have proven to be ineffective. At
present HTS nanostructures are realized by e-beam lithography in
combination with a hard mask\cite{Komissinski} and ion milling
etching or by focus ion beam patterning.\cite{Testa}\cite{Nagel} In
both cases the lateral damage of the structures is quite severe due
to structural and chemical changes mostly related to oxygen out
diffusion and/or Ga implantation. This fact puts severe limits to
the smallest HTS Josephson tunnel junction size that can still
retain its superconducting properties. In as-fabricated 200 nm wide
tunnel junctions the superconductive properties are lost during the
fabrication process and only partially recovered after post
annealing treatment\cite{UV}. Some progress has recently been
achieved in fabricating HTS nanoSQUIDs employing
nanobridges\cite{Wu}. However because of the extremely short
coherence lengths of HTS superconductors ( of the order of 1-2 nm in
the a-b planes for YBCO) the Josephson -like effect in HTS
nanobridges is still not well understood. The nanoSQUIDs have a
narrow range of working temperature, close to the critical
temperature of the nanostructure, which removes one of the main
advantage in using HTS based nanosensors.

\begin{figure}
 \includegraphics{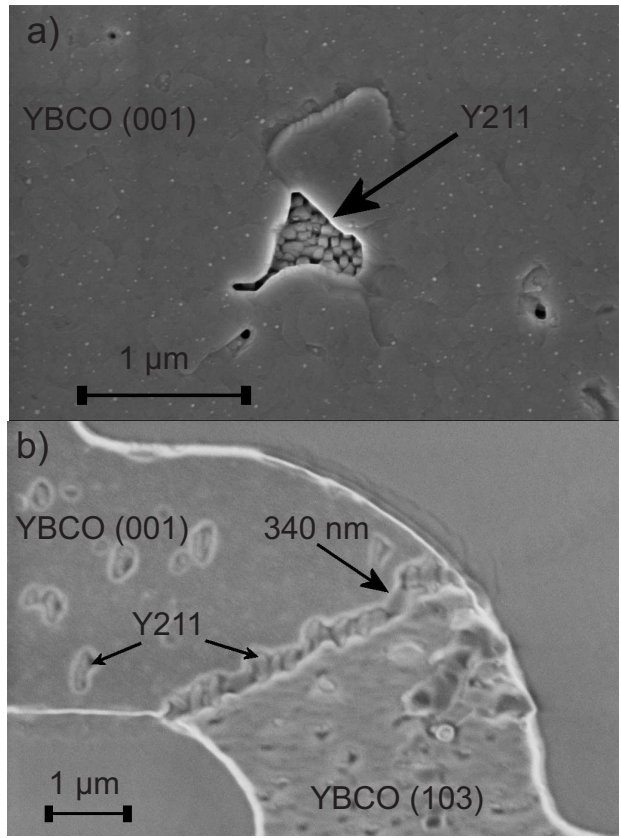}
  \caption{a) A SEM image of a Y123 (001) film grown on a MgO (110) substrate.
  A small amount of Y211 is present in the center of the image.
  b) An interface between (001) and (103) YBCO.
  The Y211 precipitate is present both in the (001) film and with a higher density at the
  interface. A 340 nm wide Y123 connection is present at the interface (marked with an
  arrow).}
  \label{Fig1}
\end{figure}

\begin{figure}
 \includegraphics{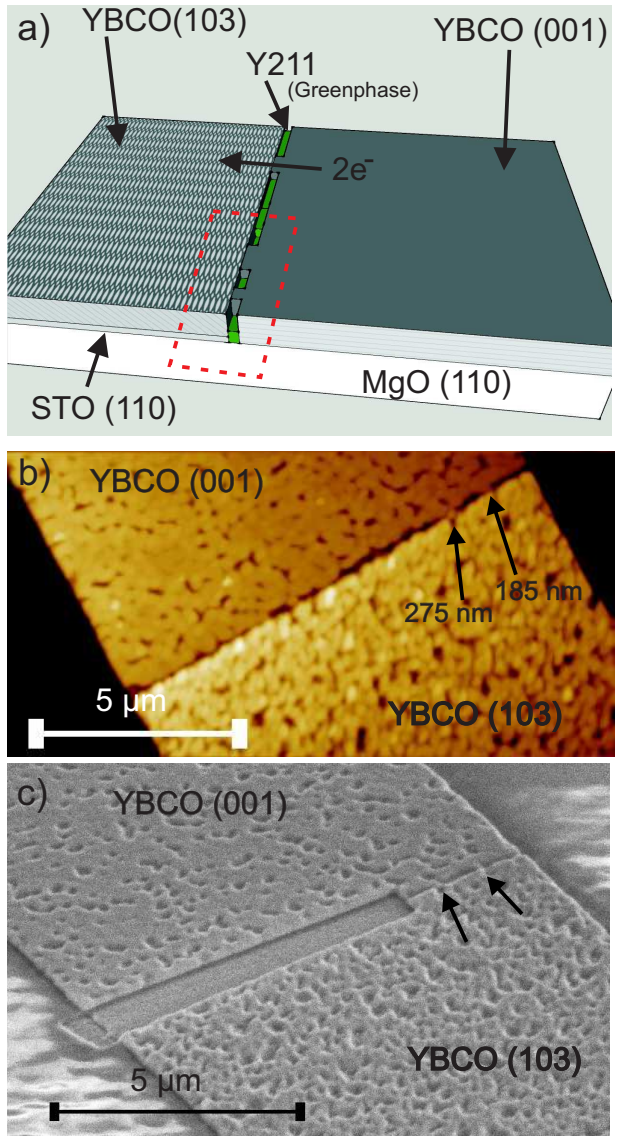}
  \caption{a) A schematic image of the (001)/(103)YBCO interface.
  3 superconducting connections are separated by segments of isolating greenphase.
  The dashed red area is cut by FIB leaving only one nanojunction. Leaving two connections
would result in a nanoSQUID. b) An AFM image of the interface before
the FIB cut. c) SEM image after the FIB, a nanoSQUID with two
nanojunctions (275 and 185 nm wide) has been created}
  \label{Fig2}
\end{figure}

In this letter we present a new method which combines conventional
nanofabrication techniques with self assembly concepts to fabricate
Josephson tunnel junctions at the nanoscale. The junctions are
realized by the biepitaxial grain boundary technique which is based
on a single layer HTS film. The main idea relies on the competition
between superconductive and insulating phases during thin film
growth. On specific substrates like MgO (110) the superconductive
YBa$_2$Cu$_3$O$_{7-\delta}$ (Y123) with [001] orientation competes
with the Y$_2$BaCuO$_5$ (Y211) insulating
phase,\cite{Green}\cite{Chiara} often referred to as green phase. In
figure \ref{Fig1} a) a SEM picture shows the two different growths.
Previous studies, made on sputtered thin films, have shown that non
optimal deposition conditions and the presence of grain
boundaries\cite{Green} strongly enhance the nucleation of the green
phase. By realizing grain boundary junctions on MgO (110), one can
expect to reach specific growth conditions where nanometer size YBCO
grain boundaries will be naturally embedded in a green phase matrix
(see figure \ref{Fig2} a)).

In what follow we will give a proof of principle of our novel soft
nanopatterning by presenting the transport properties of
nanojunctions and nanoSQUIDs. The validity of our approach has been
further confirmed by transmission electron microscopy (TEM). The TEM
provides information about the subsurface structure of the film and
also about the interface between the film and the substrate. TEM
studies are quite crucial since the superconducting thin film used
in this work are deposited by laser ablation and not by sputtering
as in previous works\cite{Green}. The different kinetics of growth
of the two deposition techniques could strongly affect the
nucleation of the greenphase and its competition with the
superconductive Y123. The transport properties of nanoscale
Josephson junctions fabricated by the soft nanopatterning technique
have also been compared with those of junctions obtained by
conventional nanofabrication procedures using an amorphous carbon
mask defined by e-beam lithography and ion milling (see
Stornaiuolo\cite{DaniellaJAP} et al. for details on the fabrication
process).

Biepitaxial grain boundary junctions are fabricated on MgO (110)
substrates using a 20 nm thick STO (110) film as seed layer. The STO
is deposited by laser ablation at an oxygen pressure of 0.2 mbar and
a deposition temperature of 680$^\circ$C. The STO film is partially
removed from the substrates by first defining an amorphous carbon
mask using e-beam lithography and finally etching the unwanted part
in an ion beam milling system\cite{Miletto}. A 100-120 nm thick YBCO
film is then deposited by laser ablation at an oxygen pressure of
0.6 mbar using a deposition temperature of 790$^\circ$C for the
conventional method and a non optimal 740$^\circ$C for the soft
pattering technique to enhance the greenphase growth. The Y123 film
grows [001] oriented on the MgO and [103] on the STO. In the films
grown at lower deposition temperature the Y123 is the dominating
phase but Y211 appears randomly on the MgO side while strongly
increasing at the MgO/STO step, see figure \ref{Fig1} b). The
fabrication of soft patterned nanometer size junctions can be
divided into two main steps:
\begin{enumerate}
\item 10 $\mu$m wide grain boundary junctions are patterned using
e-beam lithography and ion etching. The grain boundary interfaces
are then imaged using Atomic Force Microscopy (AFM) and Scanning
Electron Microscopy (SEM) to locate single or closely spaced double
nano-connections (see figure \ref{Fig2} a)).
\item Nanojunctions and nanoSQUIDs are then formed by isolating the Y123 nano connection(s) by
FIB. Since the nano devices are embedded in a green phase matrix,
their lateral sides will not be touched by the etching procedure.
Typically the distance between the nano junction and the ion beam is
at least 300-400 nm leaving the nano junction unharmed.
\end{enumerate}

\noindent Figure \ref{Fig2}b) shows an AFM image of a grain boundary
interface before, and figure \ref{Fig2}c shows a SEM image after the
FIB cut. In the final structure two junctions, 275 nm and 185 nm
wide respectively, have been preserved to form a nanoSQUID with a
loop area of the order of 0.1 $\mu$m$^2$.

Electrical measurements were performed both when the interfaces were
still 10 $\mu$m wide, and after the FIB procedure used to isolate
nanojunctions or nanoSQUIDs. The 10 $\mu$m wide junctions were also
measured before and after acquiring an image using an Ion Ga-source
with the same energy as the one used to fabricate the nanodevices.
Since the current-voltage characteristic (IVC) only changed a few
percent we concluded that the properties of the nanojunctions
forming the parallel array were not affected by the FIB cutting
procedure.

\begin{figure}
 \includegraphics{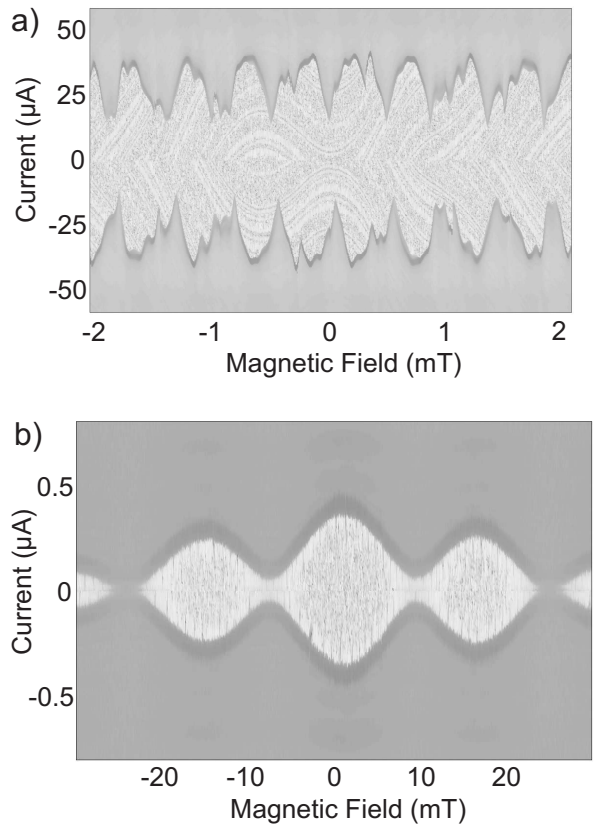}
  \caption{a) I vs B for the same junction before the FIB cut and b) after the
cut. The grey scale represents the logarithmic conductance, the
bright area (high conductance) corresponds to the super current. The
dark region (low conductance) show the critical current (I$_C$)
where the junction switch to the resistive branch.}
  \label{Fig3}
\end{figure}

For all samples we have recorded the critical current (I$_C$)
dependence on the external magnetic field (B). This measurement is
an important tool to get information about the uniformity of the
critical current distribution across the grain boundary. All the
electrical measurements were carried out at 280 mK and 34 K using an
Oxford Instruments Heliox VL $^3$He system. We have characterized a
number of single nanojunctions and nanoSQUIDs. Figure \ref{Fig3} a)
shows the modulation pattern for the 10 $\mu$m wide interface before
the FIB cut and figure \ref{Fig3} b) after a single nanojunction is
isolated by the FIB procedure. The behavior of the $I_C$ vs B
pattern before the cut shows quick and irregular modulation and
resembles that of an array of parallel connections\cite{Likarev}.
After the cut, the magnetic pattern follows an almost ideal
Fraunhofer like behavior as expected for a single Josephson
junction\cite{Tinkham}. From the period of the magnetic modulation
($\Delta$B) of the Josephson current one can extract a value for the
junction width and compare it with that estimated by AFM/SEM
pictures. Following Rosenthal and coworkers\cite{w2} the expression
for $\Delta$B which includes demagnetization effects
is\cite{note_Rosenthal1}
$\Delta$B=$\Phi_0$t/[$1.2w^2(\lambda_{eff}+\lambda_{ab} +d)]$. Here
$\Phi_0$ is the magnetic flux quantum, $\lambda$ is the London
penetration depth, t is the thickness of the film, d and w are the
length and width of the junction respectively. We have applied this
formula to a number of single junctions and the values were in good
agreement with the ones extracted from AFM/SEM pictures differing at
most by a factor less than 40\%. From the magnetic pattern of figure
\ref{Fig3}b we got w $\sim$ 100 nm which explains why this
nanojunction was not distinctively seen in the AFM picture.

\begin{figure}
 \includegraphics{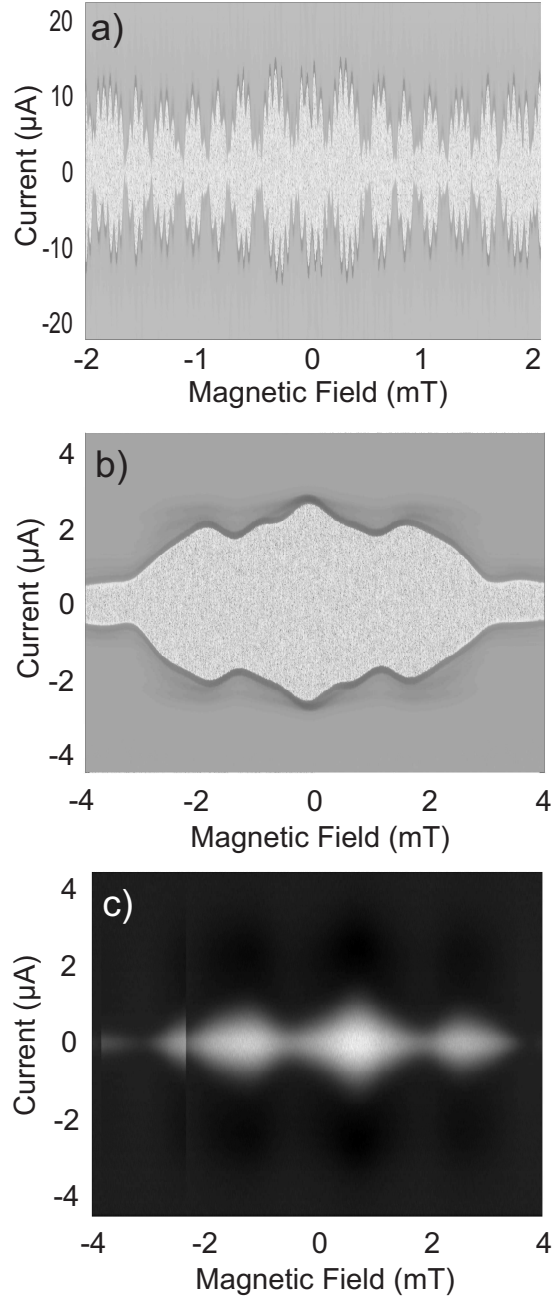}
  \caption{I vs B for the nanoSQUID shown in figure \ref{Fig2}, a) before and b) after the FIB cut.
  The measurements where performed at 300 and 280 mK respectively. c) I vs B for the same nanoSQUID at
  34 K. The amplitude of the second harmonic components decrease faster with increasing temperature than the
  first harmonic components.\cite{tobias}\cite{Löfwander} The hump structure was almost completely suppressed in this measurement.
  This further suggests that the unconventional shape of the I$_C$ vs B can
be explained by including a second harmonic component in the CPR
leading to an h/4e oscillation period (here h is Planck's constant
and e is the electron charge).}
  \label{Fig4}
\end{figure}

Figures 4 a) and b) show the I$_C$ vs B measurements for a
nanoSQUID, before and after the cut respectively. When the structure
is cut down to leave two connections separated by an area of
insulating Y211, the I$_C$ vs B clearly shows a SQUID like pattern.
Two different modulations are distinguishable; a faster one, due to
the magnetic flux linked to the SQUID loop and a slower modulation
originating from the I$_C$ vs B of two junctions. Because of the
extremely small SQUID loop area (less than 0.1 $\mu$m$^2$) even
though the junctions' widths are on the nanoscale they cannot be
considered point-like. In this case the slower modulation has a
Fraunhofer like pattern that acts as an envelope over the faster
SQUID modulation\cite{Barone}. A hump can be noticed on the side of
the inner lobes. This can be accounted for by considering that the
current-phase relation (CPR) of Josephson junctions with a d-wave
order parameter is unconventional and will have a significant second
harmonic component.\cite{tobias}\cite{Löfwander} Figure \ref{Fig4}
c) shows the I$_C$ vs B dependence of the same nanoSQUID at 34 K.
The increase in the modulation depth with temperature can probably
be attributed to a different temperature dependence of the two
components (first and second harmonics) in the current phase
relation.\cite{tobias}\cite{Löfwander} Flux trapping at the
greenphase spots, in the wide electrodes, makes magnetic field
measurements at temperatures higher than 40 K quite challenging.
This can be improved by optimizing the geometry of the
electrodes.\cite{Castellanos}\cite{Dantsker} However since the
Josephson current has an almost linear temperature dependence we
expect working nanoSQUIDs at temperatures close to 77 K.

\begin{figure}
 \includegraphics{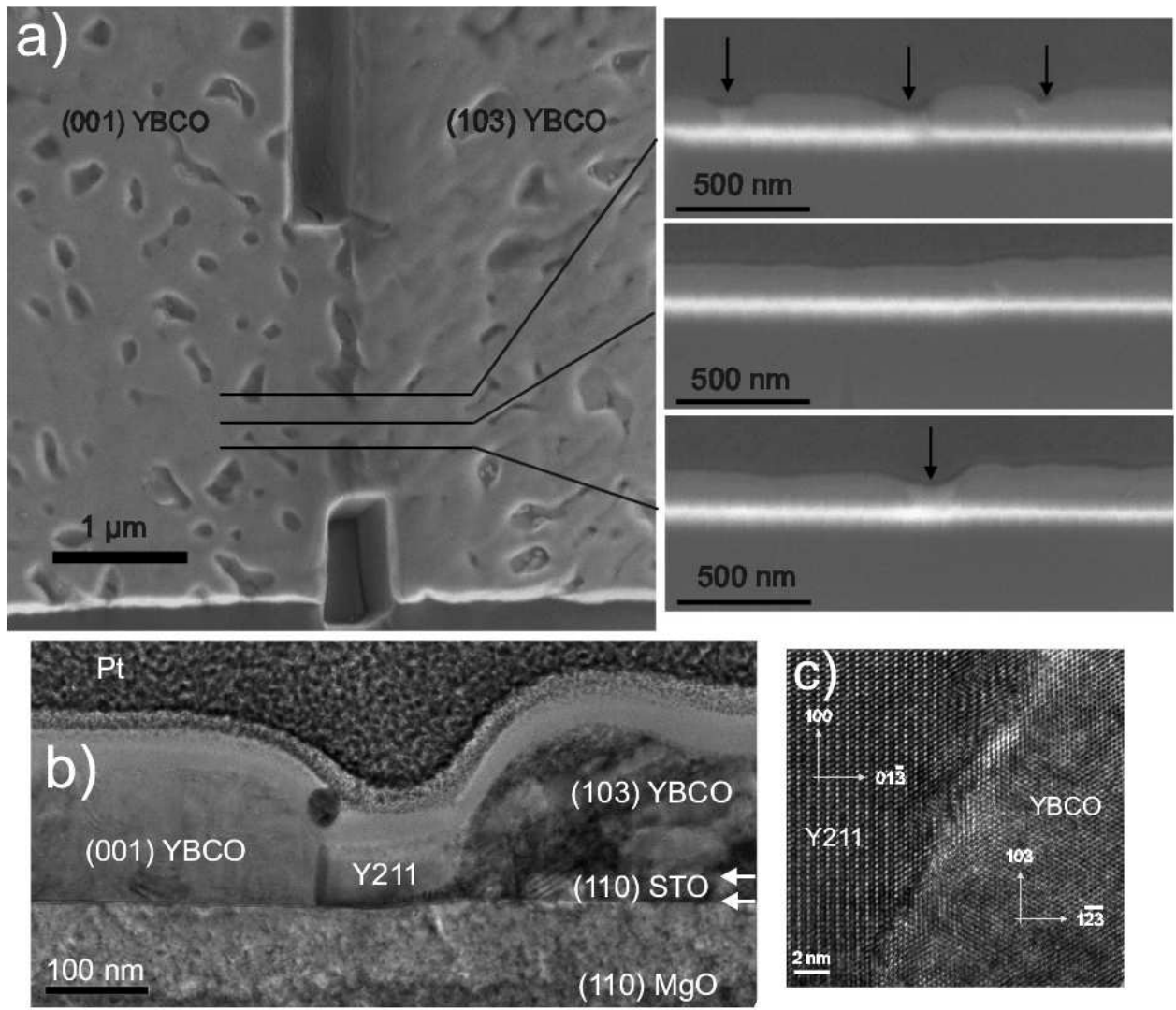}
  \caption{a) Left panel shows a SEM image of a nanoSQUID
  structure. The right panel shows three cross SEM images recorded while the sample was
thinned for TEM using FIB; the arrows mark the Y211 precipitate. b)
Low TEM magnification of the grain boundary region with Y211
precipitates. The dark contrast at the interface originates from the
specimen preparation where Pt particles in the protection layer is
pushed towards the substrate interface. Subsequent Ar ion milling
reduced the effect, revealing more details of the interface between
Y211 and the substrate and also the interface between the Y211 and
the (103) YBCO. Close examination of high resolution images (not
shown) demonstrated that the greenphase particle was in contact with
the substrate without any YBCO separating them. c) High
magnification image of the interface between the Y211 precipitate
and the (103) YBCO recorded along the [33$\bar{1}$]$_{YBCO}$}.
  \label{Fig5}
\end{figure}

The dramatic change in magnetic field dependence of the Josephson
current after the FIB procedure gives strong indications that our
new soft nanostructuring approach works. However, It do not allow to
distinguish, in an unambiguous way, between a grain boundary
consisting of an array of parallel nanochannels and a single grain
junction with a highly nonuniform critical current distribution.
This would be a possible scenario, for example, if the nucleation of
greenphase is not on the bare MgO(110) substrate but on a thin
superconductive layer due to overgrowth. To confirm that we are
probing an array of YBCO nanochannels embedded in a insulating
greenphase matrix it is important to show that the greenphase
particles extends all the way from the substrate/film interface up
to the film surface. This information cannot be extracted using SEM.
We have used high resolution TEM to reveal the interfacial
microstructure with high spatial resolution in combination with
electron diffraction to confirm the crystal structure of
Y$_2$BaCuO$_5$. The TEM specimen were prepared using a combined
FIB/SEM that allowed site specific extraction of grain boundary
junctions. Here we show the results for a nanoSQUID with
approximately the same dimensions as that of figure \ref{Fig4}
showing the same Josephson phenomenology. Cross section SEM images
were recorded during the specimen preparation, see figure \ref{Fig5}
a). A greenphase particle adjacent to a YBCO nanochannel is seen in
figure \ref{Fig5} b). It should be noted that the particle had
nucleated at the film/substrate interface and extended all the way
up to the film surface. Electron diffraction and Fourier transforms
of high resolution TEM images showed that the particle was of the
Y$_2$BaCuO$_5$ phase. This was in accordance with the energy
dispersive x-ray (EDX) analysis.

\begin{figure}
 \includegraphics{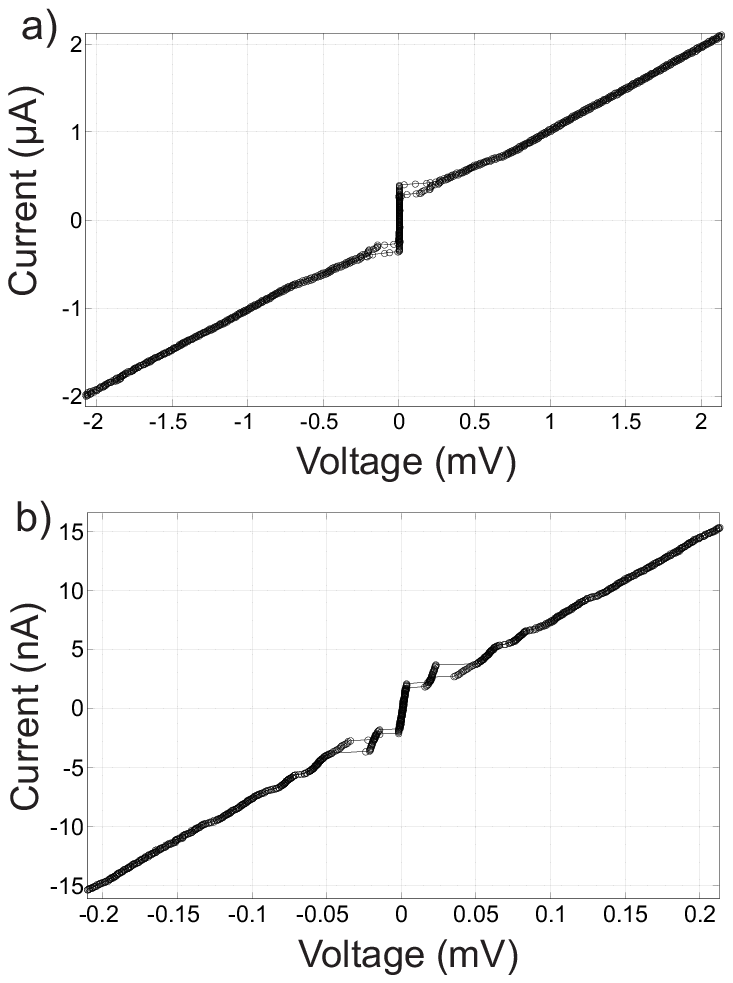}
  \caption{a) Current voltage characteristics (IVC) for a sample fabricated by soft nanopatterning with a width of 100 nm.
  b) IVC for a sample fabricated with the conventional nanopatterning
  technique. Here the nominal junction width is 300 nm, the effective width (w$_{eff}$) extracted from the modulation pattern is $\approx$60 nm.
  The presence of a second switch is due to the specific design, with two Josephson junctions in series employed to study charging effects.
  }
  \label{Fig6}
\end{figure}

It is now important to compare the transport properties of nanoscale
biepitaxial Josephson junctions made by soft nanopatterning with
those fabricated with conventional methods. Figure \ref{Fig6} a)
shows the current voltage characteristic (IVC) of a nanojunction
made by soft nanopatterning. The curve follows the Resistively
Capacitively Shunted Junction (RSCJ) model. A small hysteresis is
visible and it is typical for most of the nanodevices. Because of
the extremely small dimensions of the Josephson devices, we cannot
rule out that the hysteresis is due to heating\cite{Ransley}. For
comparison figure \ref{Fig6} b) shows the IVC of a sample of similar
dimensions fabricated by conventional nanopatterning. The two curves
show a clear difference in order of magnitudes both in Josephson
current and normal resistance.

A more detailed examination of the current voltage characteristics
for conventional junctions having widths in the range 200-300 nm
gives a clear understanding of the extent of the lateral damage. For
junctions fabricated using the conventional method with nominal
width of 200 nm no Josephson effect was observed, they instead
showed a highly resistive IVC (R$_N \sim$ 100 k$\Omega$ - 100
M$\Omega$). Only junctions with widths of at least 300 nm had a
finite critical current.

From the modulation period of the $I_C$(B) patterns for the
junctions fabricated using the conventional method it is possible to
get an estimation for the effective width of the superconducting
region and in this way determine the extent of the lateral damage.
 Considering the expression:\cite{w2}\cite{note_Rosenthal2} $\Delta$B=1.84$\Phi_0/w_{eff}^2$ one can retrieve the actual width
 of the superconducting junction, w$_{eff}$. The conventionally fabricated junction shown in figure \ref{Fig6} b) modulates on a
 field scale on the order of Tesla. For these field scales flux trapping is an issue.
 The recorded magnetic pattern was quite irregular with several jumps in $I_C$. However,
 we were able to unambiguously extract a modulation period of $\approx$ 1 T which gives w$_{eff}$$\approx$60 nm.
 The nominal size of this junction is 300 nm which indicates that a significant part of the grain boundary has lost its
 superconductive properties during the milling process. This estimation together with the fact that all 200 nm wide
junctions do not show any Josephson effect, is consistent with a non
superconducting layer of more than 100 nm formed on each side of the
device, also inferred by other groups\cite{UV}. This effect clearly
limits the smallest size we can achieve with the conventional
method.

On average samples fabricated with the soft patterning method have
critical current densities (J$_C$) in the range 10$^{3}$-10$^{4}$
A/cm$^2$. Using conventional fabrication techniques we get junctions
which have J$_C$ in the range 10$^{0}$ - 10$^{3}$ A/cm$^2$. Typical
values for the specific interface resistivites, $\rho_N$=R$_N$A, are
in the range 10$^{-8}$ - 10$^{-7}$ $\Omega$cm$^2$ for soft
nanopatterning and 10$^{-7}$-10$^{-2}$ $\Omega$cm$^2$ for
conventional nanofabrication. Here R$_N$ is the normal resistance of
the nanojunction extracted from the IVC and A is the junction area.

The huge spread in J$_C$ and $\rho_N$ for junctions fabricated with
conventional nanopatterning is certainly related to the randomness
in dimension of the superconductive grain boundary which survives
the ion milling process. However this effect alone cannot account
for the order of magnitude lower J$_C$ and higher $\rho_N$ of the
conventional nanojunctions compared to those fabricated with soft
nanopatterning. Instead, it clearly indicates that the conventional
technique creates grain boundaries that are damaged and underdoped.
These facts allow us to conclude that by the soft patterning
technique we can, for the first time, get access to pristine grain
boundaries which opens new prospectives to design experiments to
study the intrinsic properties of HTS at the nanoscale.

In conclusion we have developed a new nano fabrication process to
obtain biepitaxial YBCO grain boundary Josephson junctions at the
nanoscale. The main advantage of this technique compared to
conventional nanofabrication methods is that the grain boundaries
are unharmed by the nano patterning. Electrical transport
measurements in an externally applied magnetic field of Josephson
junctions and SQUIDs reveal very clean and homogeneous grain
boundary interfaces. Such clean interfaces are crucial for
fundamental studies of superconductivity in cuprate-materials.
Moreover the accomplishment of nano junctions allow for the
realization of nanoSQUIDs for single spin detection and quantum
circuits implementing HTS grain boundary josephson junctions.

\begin{acknowledgement}
This work has been partially supported by EU STREP project MIDAS,
the Swedish Research Council (VR) under the Linnaeus Center on
Engineered Quantum Systems, the Swedish Research Council (VR) under
the project Fundamental properties of HTS studied by the quantum
dynamics of two level systems, the Knut and Alice Wallenberg
Foundation.
\end{acknowledgement}

\bibliography{achemso}

\begin{thebibliography}{00}
\bibitem{note_Rosenthal1}$\lambda_{ab}$ for a T$_c$ of 89 K is approximately
160 nm\cite{Zuev}. $\lambda_{eff}$ = 920 nm was obtained using
$\lambda_{c}$ = 2 $\mu$m\cite{Homes} and geometrical
considerations\cite{Jesper} for a grain boundary angle of 40$^\circ$
\bibitem{note_Rosenthal2} Due to the different geometry of the electrodes used for the junctions fabricated
with conventional nanofabrication techniques, we need to use a
different expresion for the modulation period valid when
$\lambda^2/t>w$.
\end{thebibliography}

\end{document}